\documentclass[aps,prl,twocolumn,showpacs,psfig,superscriptaddress,longbibliography]{revtex4-2}
\usepackage{amsfonts}
\usepackage{mathrsfs}
\usepackage{amsmath}% needed for subequations
\usepackage{color}
\usepackage{natbib}
\usepackage{textcomp}
\usepackage{graphicx}
\usepackage{bm}% bold maths
\usepackage{amssymb}

\usepackage{xspace}
\usepackage{epstopdf}
\usepackage{dcolumn}% Align table columns on decimal point
\usepackage{longtable}
\usepackage{multirow}
\usepackage[colorlinks=true, letterpaper=true, pdfstartview=FitV, linkcolor=blue, citecolor=blue, urlcolor=blue]{hyperref}
\usepackage{float}

\makeatletter

\newcommand{\Rmnum}[1]{\expandafter\@slowromancap\romannumeral #1@}
\makeatother
\begin{document}

\title{Enhanced spin-orbit coupling and orbital moment in ferromagnets by electron correlations}

\author{Ze Liu}
\affiliation{Kavli Institute for Theoretical Sciences, and CAS Center for Excellence in Topological Quantum Computation, University of Chinese Academy of Sciences, Beijing 100190, China}

\author{Jing-Yang You}
\affiliation{Department of Physics, National University of Singapore, 2 Science Drive 3, Singapore 117551}

\author{Bo Gu}
 \email{gubo@ucas.ac.cn}
 \affiliation{Kavli Institute for Theoretical Sciences, and CAS Center for Excellence in Topological Quantum Computation, University of Chinese Academy of Sciences, Beijing 100190, China}
\affiliation{Physical Science Laboratory, Huairou National Comprehensive Science Center, Beijing 101400, China}

\author {Sadamichi Maekawa}
  \affiliation{Center for Emergent Matter Science, RIKEN, Walo 351-0198, Japan}
  \affiliation{Kavli Institute for Theoretical Sciences, and CAS Center for Excellence in Topological Quantum Computation, University of Chinese Academy of Sciences, Beijing 100190, China}

\author{Gang Su}
\email{gsu@ucas.ac.cn}
\affiliation{Kavli Institute for Theoretical Sciences, and CAS Center for Excellence in Topological Quantum Computation, University of Chinese Academy of Sciences, Beijing 100190, China}
\affiliation{Physical Science Laboratory, Huairou National Comprehensive Science Center, Beijing 101400, China}
\affiliation{School of Physical Sciences, University of Chinese Academy of Sciences, Beijing 100049, China}

\begin{abstract}
In atomic physics, the Hund rule says that the largest spin and orbital state is realized due to the interplay of the spin-orbit coupling (SOC) and the Coulomb interactions. Here, we show that in ferromagnetic solids the effective SOC and the orbital magnetic moment can be dramatically enhanced by a factor of $1/[1-(2U^\prime-U-J_H)\rho_0]$, where $U$ and $U^\prime$ are the on-site Coulomb interaction within the same oribtals and between different orbitals, respectively, $J_H$ is the Hund coupling, and $\rho_0$ is the average density of states. This factor is obtained by using the two-orbital as well as five-orbital Hubbard models with SOC. We also find that the spin polarization is more favorable than the orbital polarization, being consistent with experimental observations. This present work provides a fundamental basis for understanding the enhancements of SOC and orbital moment by Coulomb interactions in ferromagnets, which would have wide applications in spintronics.

\end{abstract}
\maketitle

%%%%%%% Main text %%%%%%%%%%%%%%%%%%%%%
%\newpage
%\section{Introduction}

{\color{blue}{\em Introduction}}---The Hund's rule in atomic physics says that the state with both the largest spin moment and the largest orbital moment is realized in an atom, required by the minimum of the Coulomb repulsive energy. The similar picture was obtained in the magnetic impurity systems. In the Anderson impurity model, the spin magnetic moment of impurities is developed due to the large on-site Coulomb interaction U~\cite{Anderson1961}. In 1964, the extended Anderson impurity model with degenerate orbitals has been studied, where the role of $U$ and the Hund coupling $J_H$ has been addressed~\cite{Moriya1965,Yosida1965}. Forty years ago, Yafet also studied the Anderson impurity model within Hartree-Fock approximation and found that the on-site Coulomb interaction of impurities can enhance the effective SOC in the spin-flip cross section~\cite{Yafet1971}. Later, Fert and Jaoul applied this result to study the anomalous Hall effect due to magnetic impurities~\cite{Fert1972}. The relation between the on-site Coulomb interaction $U$ and the effective spin-orbit coupling (SOC) in magnetic impurity systems has also been discussed by the density functional theory (DFT) calculations~\cite{Guo2009} and the quantum Monte Carlo simulations~\cite{Gu2010}.

In these years, one of the fast developing areas in condensed matter physics is spintronics~\cite{Zutic2004,Maekawa2006}. It aims to manipulate the spin rather than the charge degree of freedom of electrons to design the next-generation electronic devices with small size, faster calculating ability, and lower energy consumption. SOC, as one of the key ingredients in spintronics, is related to many significant physical phenomena and novel matter~\cite{Soumyanarayanan2016}. In addition to the magnetic anisotropy~\cite{Zutic2004,You2020}, SOC plays an important role in the phenomena such as anomalous Hall effect~\cite{Nagaosa2010}, spin Hall effect associated with the spin-charge conversion~\cite{Dyakonov1971,Hirsch1999,Kato2004,Sinova2015}, topological insulators~\cite{Fu2007,Moore2007,Hasan2010,Qi2011,You2019a}, skymions ~\cite{Muhlbauer2009,Yu2010,Nagaosa2013} and so on. To design better spintronic devices, a large SOC is usually required. As SOC is a relativistic effect in quantum mechanics, it is often small in many materials. A key issue is what factors can affect the magnitude of the SOC in solids.

On the other hand, the orbital moment in the FeCo nanogranules was experimentally shown to be about three times larger than that in bulk FeCo, as a result of the enhanced Coulomb interaction in the FeCo/insulator interface~\cite{Ogata2017}, because the Coulomb interaction in the FeCo/insulator interface is expected to be larger than that in the ferromagnetic FeCo bulk. In addition, a large Coulomb interaction up to 10 eV was discussed in Fe thin films in the experiment~\cite{Gotter2020}. The spin polarization in the Hubbard model with Rashba SOC can also be enhanced by the on-site Coulomb interaction U~\cite{Riera2013}. Recently, in the two-dimensional magnetic topological insulators PdBr$_3$ and PtBr$_3$, the DFT calculations show that the band gap and the SOC can be strongly enhanced by the Coulomb interaction ~\cite{You2019}.

Inspired by recent experimental and numerical results on the enhanced SOC due to the Coulomb interaction in strongly correlated electronic systems, here we develop a theory on the relation between SOC and Coulomb interaction in ferromagnets. By a two-orbital Hubbard model with SOC, we find that the effective SOC and orbital magnetic moment in ferromagnets can be enhanced by a factor of $1/[1-(2U^\prime-U-J_H)\rho_0]$, where $U$ and $U^\prime$ are the on-site Coulomb interaction within the same oribtals and between different orbitals, respectively, $J_H$ is the Hund coupling, and $\rho_0$ is the average density of states. The same factor has also been obtained for the five-orbital Hubbard model with degenerate bands. Our theory can be viewed as the realization of Hund's rule in ferromagnets.

{\color{blue}{\em Two-orbital Hubbard model with SOC}}---Let us consider a two-orbital Hubbard model, where only a pair of orbitals with opposite orbital magnetic quantum numbers $m$ (-1 and 1, or -2 and 2) are considered. Thus, the Hamiltonian can be written as
\begin{equation} \label{1}
\begin{aligned}
H=&\sum_{\mathbf k, m, \sigma} \epsilon_{\mathbf k m \sigma} n_{\mathbf k m \sigma}+U \sum_{i, m} n_{i m \uparrow} n_{i m \downarrow}\\
&+U^{\prime} \sum_{i, \sigma, \sigma^{\prime}} n_{i m \sigma} n_{i \bar{m} \sigma^{\prime}}-J_H \sum_{i, \sigma} n_{i m \sigma} n_{i \bar{m} \sigma},
\end{aligned}
\end{equation}
where $\epsilon_{\mathbf{k}m \sigma}$ is the energy of electron with wave vector $\mathbf k$, orbital $m$, and spin $\sigma$ $(\uparrow ,\downarrow)$~\cite{Kaplan1983}, $U$ and $U^\prime$ are the on-site Cuolomb repulsion within the orbital $m$ and between different orbitals $m$ and $m^\prime$, respectively, $J_H$ is the Hund coupling, and $n_{\mathbf{k}m \sigma}$($n_{im \sigma}$) represents the particle number with wave vector $\mathbf k$ (site index $i$), orbital $m$ and spin $\sigma$. For simplicity, we consider four degenerate energy bands, which are lifted by an external magnetic field $h$ and the Ising-type SOC~\cite{Fert1972}:
 \begin{equation} \label{2}
 \epsilon_{\mathbf k m \sigma}=\epsilon_{\mathbf k}-\sigma \mu_{B} h-\frac{1}{2} \sigma \lambda_{s o} m,
  \end{equation}
  where $\lambda_{so}$ is the SOC constant, $\epsilon_{\mathbf{k}}$ is the electron energy without external magnetic filed and SOC. Using the Hartree-Fock approximation, we have $n_{i m \sigma} n_{i m^{\prime} \sigma^\prime} \approx\left\langle n_{i m \sigma}\right\rangle n_{i m^{\prime} \sigma^\prime}+\left\langle n_{i m^{\prime} \sigma^\prime}\right\rangle n_{i m \sigma}-\left\langle n_{i m \sigma}\right\rangle\left\langle n_{i m^{\prime} \sigma^\prime}\right\rangle$. Assuming the system is homogeneous, the occupation number $n_{im\sigma}$ is independent of lattice site $i$: $\left\langle n_{im\sigma}\right\rangle \approx \left\langle n_{ m \sigma}\right\rangle$, and through Fourier transformation $\sum_{i} n_{i m \sigma}=\sum_{\mathbf{k}} n_{\mathbf{k} m \sigma}$, the Halmiltonian in Eq.(\ref{1}) can be diagonalized as:
\begin{equation} \label{3}
H\approx \sum_{\mathbf{k},m,\sigma}\tilde{\epsilon}_{\mathbf{k}m\sigma}n_{\mathbf{k}m\sigma},
\end{equation}
with
\begin{equation} \label{13}
\begin{aligned}
\tilde{\epsilon}_{\mathbf{k} m \sigma}=&\epsilon_{\mathbf{k}}-\sigma \mu_{B} h-\frac{1}{2} \sigma \lambda_{s o} m+U\left\langle n_{m \bar{\sigma}}\right\rangle\\
&+U^{\prime}\left(\left\langle n_{\bar{m} \sigma}\right\rangle+\left\langle n_{\bar{m} \bar{\sigma}}\right\rangle\right)-J_H\left\langle n_{\bar{m} \sigma}\right\rangle.
\end{aligned}
\end{equation}
We define the spin polarization per site as $s_{z}=\mu_B(\langle n_{m \uparrow}\rangle-\langle n_{m \downarrow}\rangle+\langle n_{\bar{m} \uparrow}\rangle-\langle n_{\bar{m} \downarrow}\rangle)$, and the orbital polarization per site as $l_{z}= m\mu_B(\langle n_{m \uparrow}\rangle-\langle n_{\bar{m} \uparrow}\rangle+\langle n_{m \downarrow}\rangle-\langle n_{\bar{m} \downarrow}\rangle)$. Here we should remark that the so-defined orbital polarization from itinerant electrons on different orbitals with SOC differs from the conventional orbital moments of atoms that are usually quenched owing to the presence of the crystal fields in transition metal ferromagnets. Introduce the particle numbers of the parallel ($n_{p}$) and antiparallel ($n_{ap}$) states of the spin $\sigma$ and orbital $m$: $n_{ p}$$=$$\langle n_{{m} \uparrow}\rangle+\langle n_{\bar m \downarrow}\rangle$, $n_{a p}$$=$$\langle n_{\bar{m} \uparrow}\rangle+\langle n_{m \downarrow}\rangle$. Then the energy $\tilde E_{km \sigma}$ can be written as
\begin{equation}
\begin{aligned} \label{4}
\tilde{\epsilon}_{k m \sigma}=&\bar \epsilon-\sigma \mu_{B}\left(h+\frac{U+J_H}{4 \mu_{B}^{2}} s_{z}\right) \\
&-\frac{1}{2} m\left(\sigma \lambda_{s o}-\frac{U-2U^{\prime}+J_H}{2 \mu_{B} m^{2}} l_{z}\right).
\end{aligned}
\end{equation}
 When spin $\sigma$ and orbital $m$ are antiparallel (parallel) the energy $\bar \epsilon=(\epsilon_{k}+\frac{1}{2}Un_{a p(p)}+\frac{1}{2}U^{\prime}n_{a p(p) }+\frac{1}{2}U^{\prime}n_{p(ap)}-\frac{1}{2}J_Hn_{a p(ap) })$.

\begin{table*}
\renewcommand\arraystretch{2}
\caption{Comparison of the theoretical results among the Anderson impurity model, the one-orbital Hubbard model (Stoner model), and our two- and five-orbital Hubbard models with the spin-orbit coupling (SOC). $s_z$ and $l_z$ are the spin and orbital polarization, respectively. The instability conditions (IC) of $s_z$ and $l_z$ in these models are listed. $\lambda_{so}^{\mathrm{eff}}$ is the effective SOC affected by atomic SOC $\lambda_{so}$, the electron correlations $U$, $U^\prime$ and $J_H$, and the electron density of state $\rho$. The equations of five-orbital Hubbard model can be found in the Supplementary Information.}
\begin{tabular}{l|p{3.6cm}<{\centering}p{3.2cm}<{\centering}p{4.7cm}<{\centering}p{4.6cm}<{\centering}}
%\toprule[1pt]
\hline
%\hline
% & \multicolumn{4}{c}{Theory model}\\
  \hline
               & Anderson impurity model & One-orbital Hubbard model (Stoner) &  Two-orbital Hubbard model with SOC($m=\pm1$ or $m=\pm2$)& Five-orbital Hubbard model with SOC ($m=0$, $\pm1$, $\pm2$)\\
\hline
$s_z$   & --    & $\frac{2\mu_B^2\rho(E_F)}{1-U\rho(E_F)}h $ ~\cite{1938a}    &  $\frac{4\mu_B^2\rho_0}{1-(U+J_H)\rho_0}h$ [Eq.(\ref{6})]   & $\frac{10\mu_B^2\rho_0}{1-(U+4J_H)\rho_0}h$ [Eq.(63)]     \\
$l_z$   &  --    &   --     & $\frac{m^2\mu_B\rho_s}{1-(2U^\prime -U-J_H)\rho_0}\lambda_{so}$  [Eq.(\ref{8})]   &$\frac{\mu_B(\rho_{1s}+4\rho_{2s})}{1-(2U^\prime -U-J_H)\rho_0}\lambda_{so}$  [Eq.(78)]     \\
IC of $s_z$           & $(U+4J_H)\rho(E_F)>1$~\cite{Moriya1965,Yosida1965}      &  $U\rho (E_F)>1$ ~\cite{1938a}     & $(U+J_H)\rho_0>1$ [Eq.(\ref{7})]   & $(U+4J_H)\rho_0>1$    [Eq.(65)]   \\
IC of $l_z$         &  --    &  --      &\multicolumn{2}{c}{$(2U^\prime -U-J_H)\rho_0>1$  [Eq.(\ref{10})] }          \\
$\lambda_{so}^{\mathrm{eff}}$         & $\frac{\lambda_{at}}{1-(U-J_H)\rho(E_F)}  $  ~\cite{Yafet1971} &  --   &    \multicolumn{2}{c}{$\frac{\lambda_{so}}{1-(2U^\prime-U-J_H)\rho_0}$ [Eq.(\ref{9})]}          \\
\hline
\hline
\end{tabular}
\label{T-1}
\end{table*}

{\color{blue}{\em Spin polarization}}---
It is noted that without external magnetic field $h$ and SOC $\lambda_{so}$, the four energy bands with spin $\sigma $ ($\uparrow $ and $\downarrow$) and orbital $m$ (for example $1$ and $-1$) are degenerate, and the occupation numbers $n_{ap}=n_{p}$. In terms of the translational symmetry of the lattice system: $\langle n_{m \sigma}\rangle=\frac{1}{N} \sum_{i}\langle n_{i m \sigma}\rangle=\frac{1}{N} \sum_{\mathbf k}\langle n_{\mathbf k m \sigma}\rangle=\frac{1}{N} \sum_{\mathbf k} f(\tilde{\epsilon}_{\mathbf k m \sigma})$, where $f$ is the Fermi distribution function, the spin polarization can be written as $s_{z}=\frac{\mu_{B}}{N} \sum_{k}[f(\tilde{\epsilon}_{k m \uparrow})-f(\tilde{\epsilon}_{k m \downarrow})+f(\tilde{\epsilon}_{k \bar{m} \uparrow})-f(\tilde{\epsilon}_{k \bar{m} \downarrow})]$. For the system with a paramagnetic (PM) state ($h=0$), $f(\tilde \epsilon_{\mathbf k m \sigma})$ can be expanded according to $h$, which is a small value compared to Fermi energy, and $n_{ap}=n_p$, $s_z=\mu_{B} \sum_{k}[f(\tilde{\epsilon}_{PM, k m \uparrow})-f(\tilde{\epsilon}_{PM, k m \downarrow})+f(\tilde{\epsilon}_{PM, k \bar{m} \uparrow})-f(\tilde{\epsilon}_{PM, k \bar{m} \downarrow})]=0$. Up to the linear order of $h$, the spin polarization becomes
\begin{equation}  \label{6}
s_z=\frac{4\mu_B^2\rho_0 }{1-(U+J_H)\rho_0 }h,
\end{equation}
where $\rho_{0}$$=$$\frac{1}{4}$$\int_{0}^{\infty}$$[-\frac{\partial f(E)}{\partial E}][\rho_{m \uparrow}(E)+\rho_{\bar{m} \uparrow}\left(E\right)+\rho_{m \downarrow}(E)+\rho_{\bar{m} \downarrow}(E)] d E$ is the average density of states of the four energy bands. The instability condition of the spin polarization is
\begin{equation}  \label{7}
(U+J_H)\rho _0>1.
\end{equation}
This condition can be taken as an extension of Stoner criterion in the presence of SOC in itinerant ferromagnets.

{\color{blue}{\em Orbital polarization}}---
Similarly, the orbital polarization can be expressed as $l_{z}= \mu_{B} m(\langle n_{m \uparrow}\rangle-\langle n_{\bar{m} \uparrow}\rangle+\langle n_{m \downarrow}\rangle-\langle n_{\bar{m} \downarrow}\rangle) =\frac{\mu_{B}m}{N} \sum_{k}[f(\tilde{\epsilon}_{k m \uparrow})-f(\tilde{\epsilon}_{k \bar{m} \uparrow})+f(\tilde{\epsilon}_{k m \downarrow})-f(\tilde{\epsilon}_{k \bar{m} \downarrow})]$. For the ferromagnetic (FM) state, the SOC can be regarded as a small value~\cite{Fert1972}, so $f(\tilde \epsilon_{\mathbf k m \sigma})$ can be expanded according to $\lambda _{so}$, and when $\lambda_{so}=0$, $n_{ap}=n_{p}$, the zero-order term is zero. To the linear order of $\lambda_{so}$, the orbital polarization gives
\begin{equation}  \label{8}
l_{z}=\frac{ m^{2}\mu_{B}\rho_s}{1-\left(2 U^{\prime}-U-J_H \right)\rho_0} \lambda_{s o},
\end{equation}
where $\rho_s$$=$$\frac{1}{2}$$\int_{0}^{\infty}$$[-\frac{\partial f(E)}{\partial E}][\rho_{m \uparrow}(E)+\rho_{\bar{m} \uparrow}(E)-\rho_{m \downarrow}(E)-\rho_{\bar{m} \downarrow}(E)] d E$ is the average spin polarized density of states. Then Eq.(\ref{8}) can be rewritten as $l_{z}=\mu_{B} m^{2}\rho_s \lambda_{so}^{\mathrm{eff}}$, where the effective SOC $\lambda_{so}^{\mathrm{eff}}$ is
\begin{equation} \label{9}
\lambda_{so}^{\mathrm{eff}}=\frac{\lambda_{s o}}{1-\left(2 U^{\prime}-U-J_{H}\right) \rho_0}.
\end{equation}
One may note that the orbital polarization discussed here [Eq. (\ref{8})] is totally induced by the SOC, which can be enhanced by increasing $U^\prime$ or decreasing $U$ and $J_H$, we will discuss this in detail. In the absence of the SOC, such an orbital polarization is absent according to Eq. (\ref{8}). The instability condition of orbital polarization would be:
\begin{equation}  \label{10}
(2U^\prime -U-J_H)\rho_0>1.
\end{equation}
The detailed derivation is given in the Supplementary Information.

{\color{blue}{\em Five-orbital Hubbard model with SOC}}---Our theory can be easily extended to the five-orbital Hubbard model with degenerate bands, and the detailed derivation is given in the Supplementary Information. For the five-orbital case, the instability condition of the spin polarization becomes as $(U+4J_H)\rho_{0}>1$. The same expression has been obtained for the presence of localized spin moment in the Anderson impurity model with degenerate orbitals~\cite{Moriya1965,Yosida1965}. The obtained instability condition of the orbital polarization is $(2U^\prime-U-J_H)\rho_{0}>1$, which is the same as Eq.(\ref{10}) for the two-orbital case. In the five-orbital case, the effective SOC and the orbital magnetic moment can also be enhanced by a factor of $1/[(2U^\prime-U-J_H)\rho_{0}]$, that is the same enhancement factor as in the two-orbital case.

{\color{blue}{\em Discussion}}---The comparison between our theory, the Stoner model and the Anderson impurity model is shown in Table~\ref{T-1}. It is interesting to note that the instability conditions of $s_z$ between our five-orbital Hubbard model with SOC and the Anderson impurity model are the same, while the obtained effective SOC $\lambda_{so}^{\mathrm{eff}}$ between the two models are different.
\begin{figure}[!!htbp]
	\includegraphics[width=8cm]{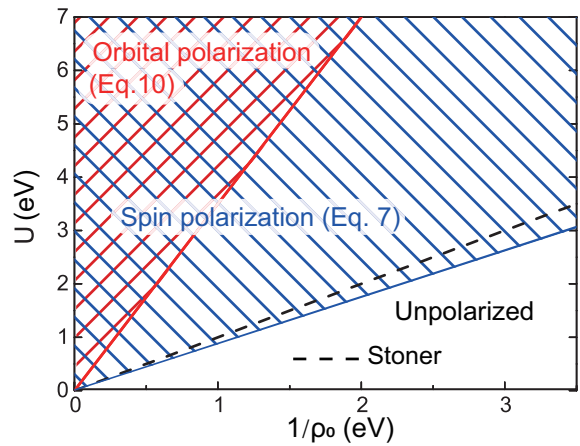}
	\caption{The phase diagram of spin and orbital spontaneous polarization as a function of the inverse average density of states and the Coulomb interaction U. The shaded area with blue solid lines represents the spin spontaneous polarization determined by Eq.(\ref{7}). The shaded area with red solid lines represents the orbital spontaneous polarization determined by Eq.(\ref{10}). The black dotted line indicates the Stoner criterion of the spin spontaneous polarization, which is obtained by the single orbital Hubbard model.}
	\label{F-2}
\end{figure}
Comparing Eqs.(\ref{7}) and (\ref{10}), which are the spin and orbital instability conditions of the two-orbital model in Table~\ref{T-1}, one may note that the condition of the orbital spontaneous polarization is more stringent than that of the spin spontaneous polarization. The phase diagram of the spin and orbital spontaneous polarizations as a function of the inverse of average density of state $1/\rho_0$ and the Coulomb interaction U obtained with Eqs. (\ref{7}) and (\ref{10}) is depicted in Fig. \ref{F-2}. Considering the relation $U=U^\prime+2J_H$ and the reasonable values of $U = 4 \sim 7$ eV in the 3d transitional metal oxides ~\cite{Maekawa2004}, for 3d electrons, $J_H=1$, $U^\prime=5$, $U=7$ eV are a set of reasonable values, for simplicity we keep the ratio $U:U^\prime:J_H=7:5:1$ in Eq.(\ref{9}), and the shaded area with blue (red) solid lines indicates the spin (orbital) spontaneous polarization. The Stoner criterion of the spin spontaneous polarization based on the single orbital Hubbard model is also plotted in Fig. \ref{F-2} for a comparison. The results show that the area of orbital spontaneous polarization is enclosed in the area of spin spontaneous polarization. In other words, it is more stringent to have the orbital spontaneous polarization, which is consistent with the fact that the orbital spontaneous polarization is rarely observed in experiments.

%\begin{figure}[!!htbp]
%	\includegraphics[width=8cm]{fig3.eps}
%	\caption{The enhancement of orbital magnetic moment $l_z$ in the FeCo nanogranules due to Coulomb interaction U. The renormalized orbital moments of FeCo bulk and FeCo nanogranules in the experiment~\cite{Ogata2017} are noted by the solid black pentagon and solid black star, respectively. The orange solid line is the result by Eq.(\ref{12}), where $\rho(E_F)$ is the density of states at Fermi energy of the FeCo interface calculated by the DFT.}
%	\label{F-3}
%\end{figure}

The relation between the electron correlations $U$, $U^\prime$ and $J_H$ and the spin polarization $s_z$ in Eq. (\ref{6}) and the orbital polarization $l_z$ in Eq. (\ref{8}) can be understood by the energy terms in Eq. (\ref{13}). For a given state with orbital $m$ and spin $\uparrow$, according to the principle of minimum energy, in order to compensate the Coulomb interaction $U$, the occupancy number $\left\langle n_{m \downarrow}\right\rangle$ will decrease, which will increase $s_z$ and decrease $l_z$. To compensate the Coulomb interaction $U^\prime$, the occupancy numbers $\left\langle n_{\bar{m} \uparrow}\right\rangle$ and $\left\langle n_{\bar{m} \downarrow}\right\rangle$ will equally decrease, which will have no effect on $s_z$ and increase $l_z$. To compensate the Hund interaction $J_H$, the occupancy number $\left\langle n_{\bar{m} \uparrow}\right\rangle$ will increase, which will increase $s_z$ and decrease $l_z$. The above argument by Eq. (\ref{13}) is consistent with the obtained enhancement factor $1/[1-(2U^\prime-U-J_H)]\rho_0$ for $l_z$ in Eq. (\ref{8}), where $U$ and $J_H$ will decrease $l_z$ and $U^\prime$ will increase $l_z$. The same argument by Eq. (\ref{13}) is also consistent with the calculated enhancement factor $1/[1-(U+J_H)]\rho_0$ for $s_z$ in Eq. (\ref{6}), where $U$ and $J_H$ will increase $s_z$ and $U^\prime$ has no effect on $s_z$.

{\color{blue}{\em Application}}---Equation (\ref{9}) shows that Coulomb interactions can enhance the effective SOC. Recently, for magnetic topological insulators PdBr$_3$ and PtBr$_3$, it is found that the energy gap increases with the increase of Coulomb interaction~\cite{You2019}. In these topological materials, the energy gap is opened due to the SOC, and the enhancement of SOC by the Coulomb interaction can be naturally obtained by Eq.(\ref{9}).
%\begin{figure}[!!htbp]
%	\includegraphics[width=8cm]{fig1.eps}
%	\caption{The enhanced energy gap and effective spin-orbit coupling due to Coulomb interaction U. The blue solid triangles and purple solid circles give the band gaps of PtBr$_3$ and PdBr$_3$, respectively, obtained by the density functional theory (DFT) calculations with different parameter U~\cite{You2019}. The blue and purple solid lines are fitted results by Eq.(\ref{9}) and Eq.(\ref{11}), where $A{\lambda_{so}}$ and $\rho_0$ are the fitting parameters.}
%	\label{F-1}
%\end{figure}
%As shown in Fig. \ref{F-1}, the blue solid triangles and purple solid circles represent the band gaps of PtBr$_3$ and PdBr$_3$, respectively, which are obtained by the DFT calculations with different parameter U~\cite{You2019}. The blue and purple solid lines are fitted results in terms of Eqs. (\ref{9}) and (\ref{11}), where A$\lambda_{so}$ and the density of state $\rho_0$ are the fitting parameters. For simplicity we use the approximation in the DFT calculation, to keep the $J_H=0$ eV, $U=U^\prime$, and study the effect of different electron correlations in the 4d and 5d transition metal compounds.
In addition, since the magnetic optical Kerr effect (MOKE) and the Faraday effect are determined by the SOC, the experimentally observed large Faraday effect in metal fluoride nanogranular films~\cite{Kobayashi2018} and the predicted large MOKE at Fe/insulator interfaces~\cite{Gu2017} can also be understood by the effect of Coulomb interaction as revealed by Eq.(\ref{9}), because the Coulomb interaction becomes important with the decreased screening effect at the interfaces.
It is noted that the Hubbard model with SOC has been extensively studied, where the SOC can induce the Dzyaloshinski-Moriya interaction and the pesudo-dipolar interaction ~\cite{Kaplan1983,Coffey1991,Shekhtman1992,Shekhtman1993,Bonesteel1993,Koshibae1993,Koshibae1993a,Viertioe1994,Yildirim1995,Tabrizi2019,Kocharian2016,Laubach2014,Koshibae1994}.

The orbital magnetic moment can also be enhanced by Coulomb interaction, as given by Eq.(\ref{8}). In the recent experiment, the orbital magnetic moment in FeCo nanogranules is observed to be three times larger than that of FeCo bulk~\cite{Ogata2017}. Using Eq.(\ref{8}), the ratio of the orbital magnetic moment without the Coulomb interaction $l_{z}(U=0)$ to the orbital magnetic moment with finite U $l_{z}(U)$ can be approximately written as:
\begin{equation}  \label{12}
\frac{l_{z}(U)}{l_{z}(0)}=\frac{1}{1-(2U^\prime-U-J_H)\rho_0}.
\end{equation}
As the Coulomb interactions can be approximately neglected in the metal bulk, and become important in the metal/insulator interfaces, $l_z(0)$ and $l_z(U)$ can represent the orbital moment of FeCo bulk and nanogranules, respectively. To reproduce the experimental ratio of orbital magnetic moment between FeCo nanogranules and bulk we may take, $l_{z}(U)/l_{z}(0)=3$, the fitted value $U = 4.4$ eV is obtained by Eq.(\ref{12}), which is reasonable for 3d transition metals. In the fitting, we use the approximation in the DFT calculation, to keep $J_H=0$ eV, $U=U^\prime$. $\rho_0 =  0.15$ (1/eV) is obtained by DFT calculation for the FeCo interface, where $\rho_0$ is approximately estimated as the density of states at Fermi level. Therefore, Eq.(\ref{12}) can be used to qualitatively explain the enhancement of orbital magnetic moment for the FeCo nanogranules in the experiment.

{\color{blue}{\em Conclusion}}---
A two-orbital Hubbard model with SOC, we show that the orbital polarization and the effective SOC in ferromagnets are enhanced by a factor of $1/[1-(2U^\prime-U-J_H)\rho_0]$, where $U$ and $U^\prime$ are the on-site Coulomb interaction within the same orbitals and between different orbitals, respectively, $J_H$ is the Hund coupling, and $\rho_0$ is the average density of states. The same factor is obtained for the five-orbital Hubbard model with degenerate bands. Our theory can be viewed as the realization of Hund's rule in ferromagnets. The theory can be applied to understand the enhanced band gap due to SOC in magnetic topological insulators, and the enhanced orbital magnetic moment in ferromagnetic nanogranules in a recent experiment. In addition, our results reveal that it is more stringent to have the orbital spontaneous polarization than the spin spontaneous polarization, which is consistent with experimental observations. As the electronic interaction in some two-dimensional (2D) systems can be controlled experimentally~\cite{Liu2021}, according to our theory, the enhanced SOC, spin and orbital magnetic moments are highly expected to be observed in these 2D systems. This present work not only provides a fundamental basis for understanding the enhancements of SOC in some magnetic materials, but also sheds light on how to get a large SOC through hybrid spintronic structures.

The authors acknowledge Q. B. Yan, Z. G. Zhu, and Z. C. Wang for many valuable discussions. This work is supported in part by the National Key R$\&$D Program of China (Grant No. 2018YFA0305800), the Strategic Priority Research Program of the Chinese Academy of Sciences (Grant No. XDB28000000), the National Natural Science Foundation of China (Grant No. 11834014), and Beijing Municipal Science and Technology Commission (Grant No. Z191100007219013). B.G. is also supported by the National Natural Science Foundation of China (Grants No. Y81Z01A1A9 and No. 12074378), the Chinese Academy of Sciences (Grants No. Y929013EA2 and No. E0EG4301X2), the University of Chinese Academy of Sciences (Grant No. 110200M208), the Strategic Priority Research Program of Chinese Academy of Sciences (Grant No. XDB33000000), and the Beijing Natural Science Foundation (Grant No. Z190011). SM is supported by JST CREST Grant (No. JPMJCR19J4 No. JPMJCR1874 and No. JPMJCR20C1) and JSPS KAKENHI (Nos. 17H02927 and 20H01865) from MEXT, Japan.

%\bibliography{ref}
%apsrev4-2.bst 2019-01-14 (MD) hand-edited version of apsrev4-1.bst
%Control: key (0)
%Control: author (8) initials jnrlst
%Control: editor formatted (1) identically to author
%Control: production of article title (0) allowed
%Control: page (0) single
%Control: year (1) truncated
%Control: production of eprint (0) enabled
%

\end{document}